\documentclass[conference, 10pt]{IEEEtran}
\usepackage{cite}

\ifCLASSINFOpdf
\else
\fi
\usepackage{subfigure}

\usepackage[cmex10]{amsmath}
\usepackage[margin=1in]{geometry}
\usepackage[pdftex]{graphicx}
\usepackage{amsfonts}
\usepackage{bm}

\usepackage{graphicx}
\usepackage{color}
\usepackage{placeins}
\usepackage{float}
\usepackage{tabularx,colortbl}

\begin{document}
%
\title{Norm-1 Regularized Consensus-based ADMM for Imaging with a Compressive Antenna}



%
\author{\IEEEauthorblockN{Juan Heredia Juesas,
Ali Molaei, Luis Tirado, William Blackwell, and Jos\'e \'A
Mart\'inez Lorenzo}}


\maketitle

\begin{abstract}
This paper presents a novel norm-one-regularized, consensus-based
imaging algorithm, based on the Alternating Direction Method of
Multipliers (ADMM). This algorithm is capable of imaging composite dielectric
and metallic targets by using limited amount of data. The distributed capabilities
of the ADMM accelerates the convergence of the imaging. Recently, a
Compressive Reflector Antenna (CRA) has been proposed as a way to
provide high-sensing-capacity with a minimum cost and
complexity in the hardware architecture. The ADMM algorithm applied to the
imaging capabilities of the Compressive Antenna (CA) outperforms current state of the art iterative reconstruction
algorithms, such as Nesterov-based methods, in terms of computational cost;
and it ultimately enables the use of a CA in quasi-real-time,
compressive sensing imaging applications.
\end{abstract}


%
\IEEEpeerreviewmaketitle

\section{Introduction}
Reducing the cost of electromagnetic sensing and imaging systems
is a necessity before they can be ubiquitously deployed as a part
or a large-scale network of sensors. Recently, a single
transceiver Compressive Antenna (CA) was proposed
as a vehicle to enhance the sensing capacity of an active imaging system, which is equivalent to maximizing the information transfer efficiency from the imaging domain and radar system; and, as a result, the cost and hardware architecture of the imaging system can be drastically reduced \cite{Martinez-Lorenzo2015}. This unique feature of CAs has triggered its use in a wide variety of applications, which include the following: 1) active imaging of metallic target at mm-wave frequencies \cite{Martinez-Lorenzo2015}; passive imaging of the physical temperature of the earth at mm-wave frequencies \cite{Molaei2016}; and active imaging of red blood cells at optical frequencies \cite{Gomez-Sousa2016}. CAs rely on the use of norm-one-regularized iterative Compressive Sensing imaging techniques (CS), such as NESTA \cite{Becker2011}, which are slow and computationally very expensive; and, ultimately, it may compromise its use in quasi-real-time imaging applications. In order to overcome these
imaging barriers, a new fully-parallelizable, consensus-based
imaging algorithm, based on the Alternating Direction Method of
Multipliers (ADMM) formulation is proposed in this paper.


\begin{figure}[htp]
\centering
        \includegraphics[scale=.315]{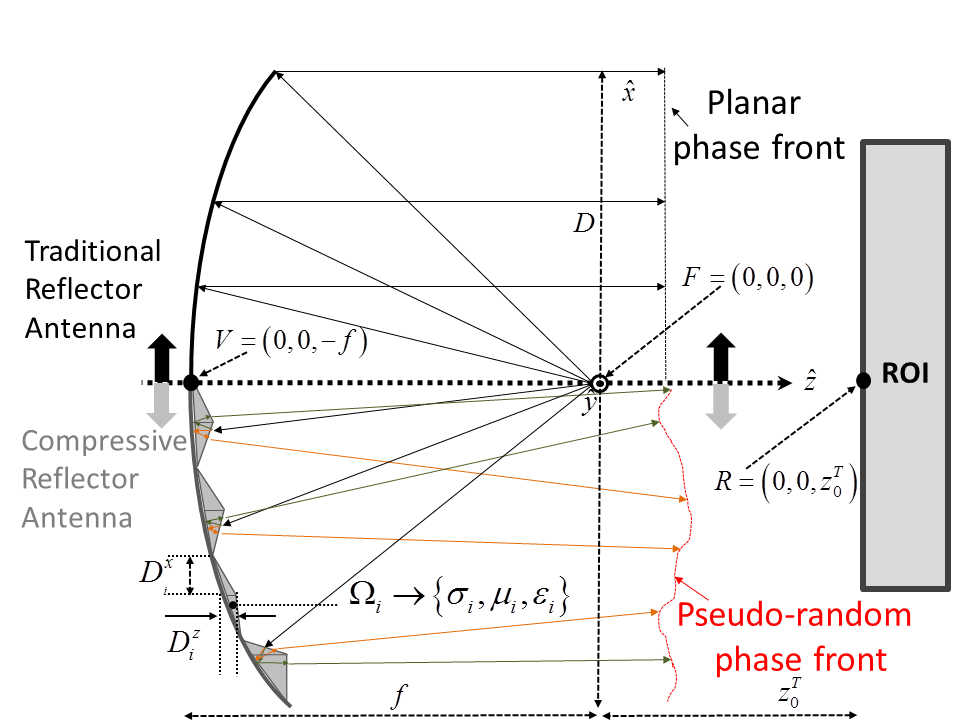}
        \caption{2D cross-section of a Traditional Reflector Antenna (${x>0}$), and Compressive Reflector Antenna (${x<0}$).}
    \label{config}
\end{figure}

\section{Compressive Reflector Antenna}

\subsection{General Overview}

The concept of operation of the CRA for sensing and imaging applications relies on two basic
principles: 1) multi-dimensional codification, generated by a customized
reflector; and 2) compressing sensing imaging, performed on the measured data.

The CRA is fabricated, as Fig. \ref{config} shows on the bottom
($x<0$), by introducing discrete scatterers, $\Omega_i$, on the
surface of a Traditional Reflector Antenna (TRA), shown on the top
($x>0$) of Fig. \ref{config}. Each scatterer $\Omega_i$ is
characterized by the electromagnetic parameters: conductivity,
permeability and permittivity, $\{\sigma_i,\mu_i,\epsilon_i\}$,
and the scatterer size $\{D_i^x,D_i^y,D_i^z\}$ in  $\{\bf \hat x,
\hat y, \hat z\}$. CRA and TRA share also many geometrical
parameters: $D$, aperture size; $f$, focal length; $h_{o}$, offset
height.

These scatterers generate a spatially coded pattern in the near
and far field of the antenna after reflecting the incident field
produced by the feeding element. When this coded pattern is
changed as a function of time, CS techniques can be used to
generate a 3D image of an object under test.
There are several techniques that may
be used for switching among different spatial coded
patterns generated by the CRA. Some of them are the following: 1) electronic beam steering by using
a focal plane array; 2) electronic beam steering by an
electronically-reconfigurable sub-reflector; 3) electronic
change of the constitutive parameters of the scatters; 4)
mechanical rotation of the reflector along the axis of the
parabola $\bf \hat{z}$; 5) mechanical rotation of a single feeding
horn or array along the axis of the parabola $\bf \hat z$.

\subsection{Sensing matrix}
For the example carried out in this paper, a mechanical rotation of the reflector along the
axis of the parabola is chosen to generate the coded pattern, so just with a single transceiver the CRA can perform the 3D imaging. This
configuration can be described as a multiple mono-static one, in
which data is collected during the scan period, $t_s$, where the
reflector is rotated $\theta_r$ degrees for $r=1,...,N_\theta$ along the axis of the parabola.
The image reconstruction, which is placed on a Region Of Interest (ROI) located $z_0^T$ meters away of the focal point of the CRA, is performed in $N_p$ pixels and the
systems uses $N_f$ frequencies. Under this configuration, the
sensing matrix $\textbf{H}\in {\mathbb{C}}^{N_t\times N_p}$ establishes a
linear relationship between the unknown complex vector $\textbf{u} \in
{\mathbb{C}}^{N_p}$ and the measured complex field data $\textbf{g} \in
{\mathbb{C}}^{N_t}$, with $N_t=N_\theta\cdot N_f$, the total number of reflector rotation
angles times the number of frequencies. This
relationship can be expressed in a matrix form as follows:
\begin{equation}
\bf{g=Hu+w},
\label{sensing_eq}
\end{equation}
where $\textbf{w}\in\mathbb{C}^{N_t}$ represents the noise collected by the
receiving antenna for a given frequency and rotation angle.

\section{ADMM formulation}
\begin{figure}[htp]
\centering
        \includegraphics[scale=.25, trim = 2mm 30mm 0mm 10mm, clip]{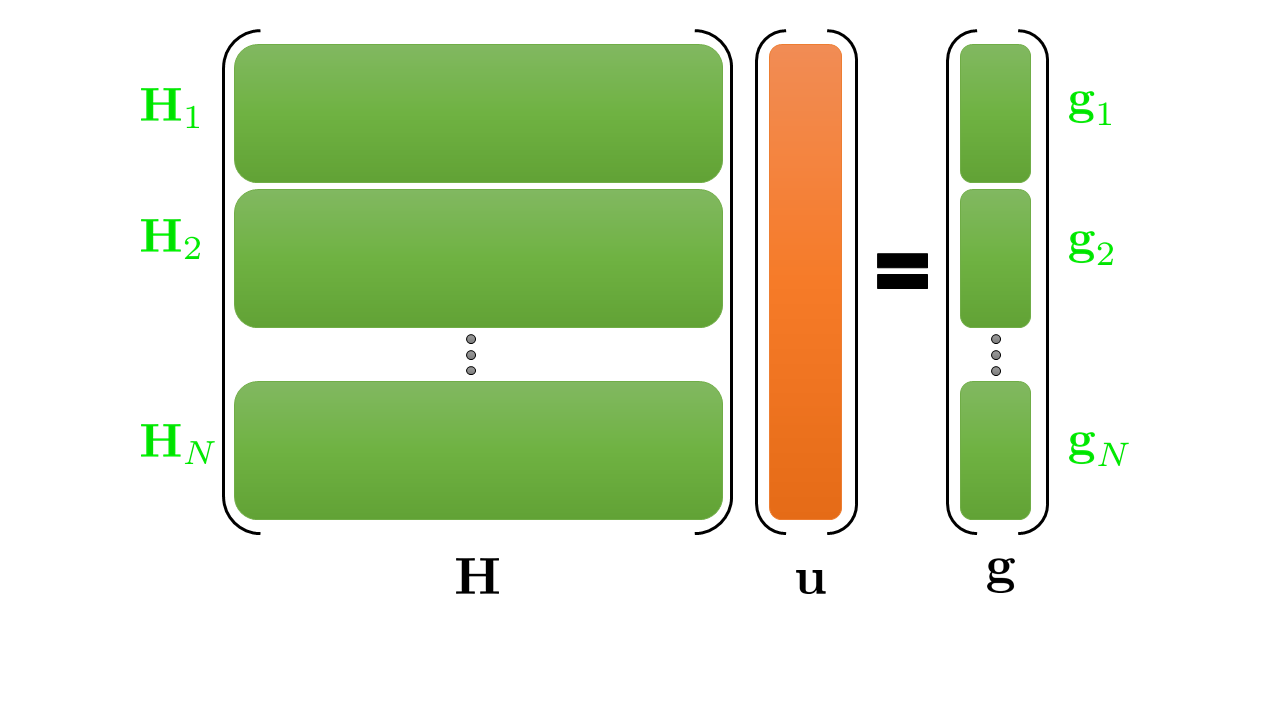}
        \caption{Division of the system by rows.}
    \label{ByRows}
\end{figure}

Equation \eqref{sensing_eq} can be solved via a novel method for optimizing convex functions called the Alternating Direction Method of Multipliers (ADMM),
\cite{boyd2011distributed, boyd2009convex}. The general
representation of an optimization problem through the ADMM takes the following form:
\begin{equation}
\left.
\begin{array}{c}
\mbox{minimize $f(\textbf{u})+g(\textbf{v})$} \\
\mbox{$\;\;\;$s.t.$\;\;\;\;\;\;\bf{Pu+Qv=c}$}%
\end{array}%
\right.
\end{equation}%
where $f$ and $g$ are convex, closed and proper functions over the unknown vectors $\textbf{u}\in{\mathbb{C}}^{n}$ and $\textbf{v}\in{\mathbb{C}}^{m}$, and the known matrices $\textbf{P}\in{\mathbb{C}}^{p \times n}$ and $\textbf{Q}\in{\mathbb{C}}^{p \times m}$ and vector $\textbf{c}\in{\mathbb{C}}^{p}$ are the ones that define the constraint. As it can be noticed, the methodology of ADMM introduces a new variable $\textbf{v}$ in order to be able to update both variables $\textbf{u}$ and $\textbf{v}$ in an \textit{alternating direction} fashion. The price to pay for this is the need to add a new constraint. A detail description
about the ADMM may be found in the references \cite{boyd2011distributed,
bauschke2011fixed, he2000alternating}.
In order to solve equation \eqref{sensing_eq} for the unknown variable $\textbf{u}$, the convex function $f(\textbf{u})=\left\Vert \bf{Hu-g}\right\Vert _{2}^{2}$ is minimized in conjuntion with the norm 1 regularized $g(\textbf{v})=\lambda \left\Vert \textbf{v}\right\Vert _{1}$; as a result, the ADMM problem to minimize takes the \textit{lasso} form and is formulated as follows:
\begin{equation}
\left.
\begin{array}{cc}
\mbox{minimize} & \frac{1}{2}\left\Vert \bf{Hu-g}\right\Vert _{2}^{2}+\lambda \left\Vert \textbf{v}\right\Vert _{1} \\
\mbox{s.t.} & \bf{u-v=0},%
\end{array}
\right.
\label{original}
\end{equation}
where $\bf{P=I}$, $\bf{Q=-I}$ and $\bf{c=0}$ enforces that the variables $\textbf{u}$ and $\textbf{v}$ are equal.
This problem can be solved in a distributed fashion, by splitting the original matrix $\bf{H}$ and the vector $\bf{g}$ into $N$ submatrices $\textbf{H}_i$ --by rows-- and $N$ sub-vectors $\textbf{g}_i$, as shown in Fig. \ref{ByRows}. Additionally, it is possible to define $N$ different variables $\textbf{u}_i$; so that the equation (\ref{original}) turns into
\begin{equation}
\left.
\begin{array}{cc}
\mbox{minimize } & \frac{1}{2}\sum\limits_{i=1}^{N}\left\Vert
\textbf{H}_{i}\textbf{u}_i-\textbf{g}_{i}\right\Vert _{2}^{2}+\lambda\left\Vert \textbf{v}\right\Vert _{1} \\
\mbox{s.t.} & \textbf{u}_i=\textbf{v},\;\;\forall i=1,...,N.%
\end{array}%
\right.
\label{sum_N}
\end{equation}%

Equation (\ref{sum_N}) is solved as $N$ different problems. The
variable $\bf{v}$ works as a \textit{consensus} variable, imposing
the agreement between all the variables $\textbf{u}_i$. See for
example \cite{degroot1974reaching, erseghe2011fast,
mota2012distributed, mota2013d, forero2010consensus}. The
augmented Lagrangian function for this problem is of the following
form:
\begin{eqnarray}
L\left( \textbf{u}_{i},\textbf{v},\textbf{s}_{i}\right)  =\frac{1}{2}\sum\limits_{i=1}^{N}\left\Vert
\textbf{H}_{i}\textbf{u}_{i}-\textbf{g}_{i}\right\Vert _{2}^{2}+\lambda\left\Vert \textbf{v}\right\Vert _{1}+  \notag \\
+\frac{\rho }{2}\sum\limits_{i=1}^{N}\left\Vert \textbf{u}_{i}-\textbf{v}+\textbf{s}_{i}\right\Vert _{2}^{2}-\frac{\rho }{2}\sum%
\limits_{i=1}^{N}\left\Vert \textbf{s}_{i}\right\Vert _{2}^{2}.
\end{eqnarray}
where $\textbf{s}_i$ is the dual variable for each constraint $i$, and $\rho $ is the augmented parameter that enforces the convexity of the function.
This problem can be solved by the following iterative scheme:
\begin{eqnarray}
\label{x_solution}
\textbf{u}_{i}^{k+1} &=&\left( \textbf{H}_{i}^{\ast }\textbf{H}_{i}+\rho \textbf{I}\right) ^{-1}\left(\textbf{H}_{i}^{\ast }\textbf{g}_{i}+\rho \left(\textbf{v}^{k}-\textbf{s}_{i}^{k}\right) \right) , \\
\textbf{v}^{k+1} &=&\mathbf{S}_{\frac{\lambda}{\rho N}}\left( \bar{\textbf{u}}^{k+1}+\bar{\textbf{s}}^{k}\right) , \\
\textbf{s}_{i}^{k+1} &=&\textbf{s}_{i}^{k}+\textbf{u}_{i}^{k+1}-\textbf{v}^{k+1} ,
\end{eqnarray}%
where $\mathbf{S}_{\kappa }\left( a\right) $ is the soft thresholding
operator \cite{beck2009fast, bredies2008linear} interpreted elementwise, defined as follows:
\begin{equation}
\mathbf{S}_{\kappa }\left( a\right) =\left\{
\begin{array}{c}
a-\kappa \ \ \ \ \ a>\kappa , \\
0\ \ \ \ \ \ \ \ \ \left\vert a\right\vert \leq \kappa , \\
a+\kappa \ \ \ \ a<-\kappa ,%
\end{array}%
\right.
\end{equation}%
$\bar{\textbf{u}}$ and $\bar{\textbf{s}}$
are the mean of $\textbf{u}_{i}$ and $\textbf{s}_{i}$, respectively, for all $i$. The variable $\bf{v}$ is used to impose the consensus, by using all the independent solutions
 $\textbf{u}_{i}$ and $\textbf{s}_{i}$.
The term $\left( \textbf{H}_{i}^{\ast }\textbf{H}_{i}+\rho \textbf{I}\right) ^{-1}$ requires the inversion of a $N_t\times N_t$ matrix, which is computationally expensive. However,
the \textit{matrix inversion lemma} \cite{woodbury1950inverting} can be applied in order to perform $N$ inversions of matrices of reduced size $\frac{N_p}{N}\times \frac{N_p}{N}$, as equation \eqref{inversion_lemma} shows:
\begin{equation}
\label{inversion_lemma}
\left( \textbf{H}_{i}^{\ast }\textbf{H}_{i}+\rho \textbf{I}_{N_t}\right) ^{-1}=\frac{\textbf{I}_{N_t}}{\rho }-\frac{%
\textbf{H}_{i}^{\ast }}{\rho ^{2}}\left( \textbf{I}_{\frac{N_p}{N}}+\frac{\textbf{H}_{i}\textbf{H}_{i}^{\ast }}{\rho }%
\right) ^{-1}\textbf{H}_{i}
\end{equation}
where $\textbf{I}_{N_t}$ and $\textbf{I}_{\frac{N_p}{N}}$ indicates the identity matrices of sizes $N_t$ and ${\frac{N_p}{N}}$, respectively.

\begin{table}[htp]
\centering \caption{Parameters for the numerical
example.}\label{tab:config}
\setlength{\extrarowheight}{1.5pt}
\begin{tabular}{|l|l||l|l|}
\hline
\multicolumn{1}{|c|}{\textbf{PARAM.}} & \multicolumn{1}{c||}{\textbf{CONFIG.}} & \multicolumn{1}{c|}{\textbf{PARAM.}} & \multicolumn{1}{c|}{\textbf{CONFIG.}} \\
\hline
\multicolumn{1}{|c|}{$\lambda _{c}$} & \multicolumn{1}{c||}{$5\cdot 10^{-3}m$} & \multicolumn{1}{c|}{$\theta _{r}$} & \multicolumn{1}{c|}{$90^{\circ }$} \\
\hline
\multicolumn{1}{|c|}{$D$} & \multicolumn{1}{c||}{$200\lambda _{c}$} & \multicolumn{1}{c|}{$N_{f}$} & \multicolumn{1}{c|}{$3$} \\
\hline
\multicolumn{1}{|c|}{$\left\langle D^{x}\right\rangle =\left\langle D^{y}\right\rangle $} & \multicolumn{1}{c||}{$1.5\lambda _{c}$} & \multicolumn{1}{c|}{$N_{p}$} & \multicolumn{1}{c|}{$25000$} \\
\hline
\multicolumn{1}{|c|}{$D_{i}^{z}$} & \multicolumn{1}{c||}{$U\left( \pm0.54\lambda _{c}\right) $} & \multicolumn{1}{c|}{$z_{0}^{T}$} & \multicolumn{1}{c|}{$195\lambda _{c}$} \\
\hline
\multicolumn{1}{|c|}{$f$} & \multicolumn{1}{c||}{$200\lambda _{c}$} & \multicolumn{1}{c|}{$\Delta x_{0}^{T}$} & \multicolumn{1}{c|}{$36\lambda_{c}$} \\
\hline
\multicolumn{1}{|c|}{$h_{0}$} & \multicolumn{1}{c||}{$0\lambda _{c}$} & \multicolumn{1}{c|}{$\Delta y_{0}^{T}$} & \multicolumn{1}{c|}{$36\lambda_{c}$} \\
\hline
\multicolumn{1}{|c|}{$N_{t}$} & \multicolumn{1}{c||}{$93$} & \multicolumn{1}{c|}{$\Delta z_{0}^{T}$} & \multicolumn{1}{c|}{$7.5\lambda_{c}$} \\
\hline
\multicolumn{1}{|c|}{$N_{\theta }$} & \multicolumn{1}{c||}{$31$} & \multicolumn{1}{c|}{$l$} & \multicolumn{1}{c|}{$1.5\lambda _{c}$} \\
\hline
\end{tabular}
\end{table}

\section{Numerical Results}
The performance of the CRA is evaluated in a millimeter-wave
imaging application. The parameters used for the numerical
simulation are shown in Table \ref{tab:config}, as defined in
\cite{Martinez-Lorenzo2015}. The total number of measurements
used for the reconstruction is given by the number of angles times
the number of frequencies, which is equal to 93. The center
frequency of the system is 60GHz, and it has a bandwidth of 6GHz.
For this example, each scatter $\Omega_i$ is considered as a
Perfect Electric Conductor (PEC), so $\sigma_i=\sigma_{PEC}$ and
the CRA is discretized into triangular patches, as described in
\cite{Meana2010}. These triangles are characterized by an averaged
size of $<D^x>$ and $<D^y>$ in $\bf \hat x$ and $\bf \hat y$
dimensions, respectively. The scatterer size $D_i^z$ of each
triangle in $\bf \hat z$ is modeled as a uniform random variable.
The parameter $\lambda_c$ is the wavelength at the center
frequency. The imaging ROI is located $z^T_0$ away from the focal
point of the CRA; and it encloses a volume determined by the
following dimensions: $\Delta x^T_0 $, $\Delta y^T_0$ and $\Delta
z^T_0$ in $\bf \hat x$, $\bf \hat y$ and $\bf \hat z$ dimensions.
The ROI is discretized into cubes of side length $l$.

\begin{figure}
     \centering
     \subfigure[]{
          \label{fig:CRA-SAR}
          \includegraphics[scale=.40, trim = 72mm 10mm 30mm 15mm, clip]{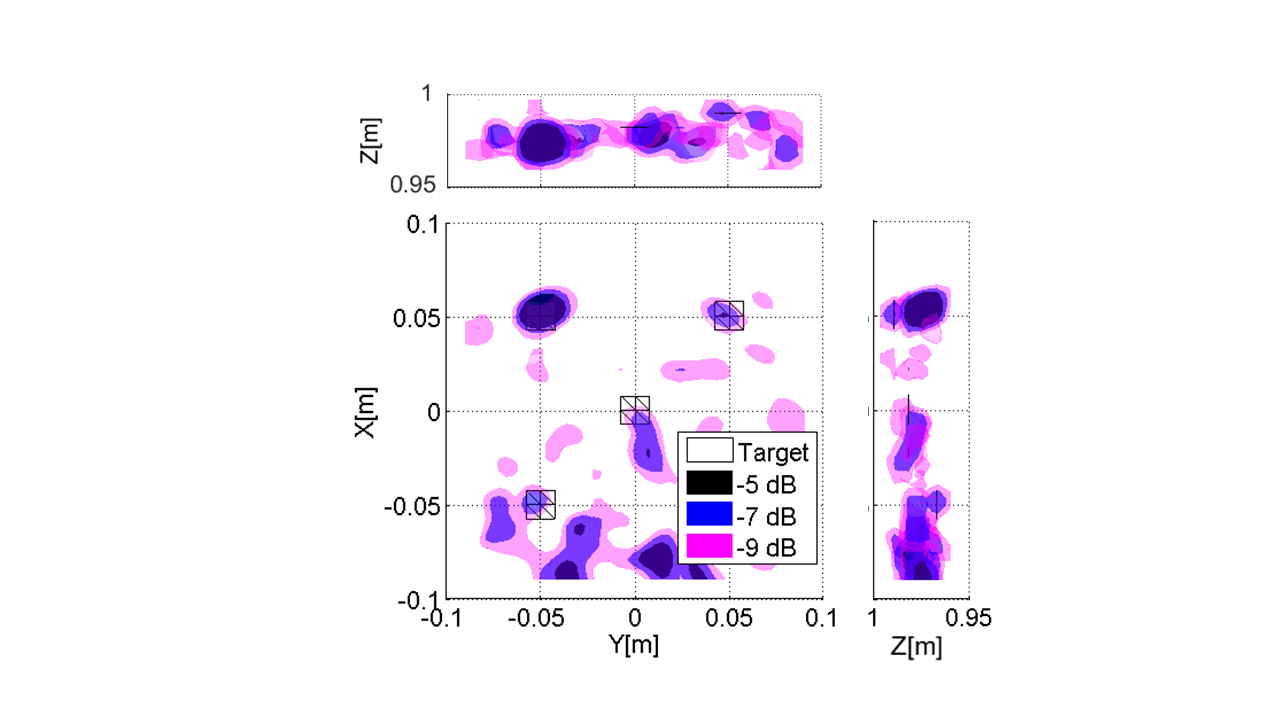}}
     \subfigure[]{
          \label{fig:CRA-NESTA}
          \includegraphics[scale=.40, trim = 72mm 10mm 30mm 15mm, clip]{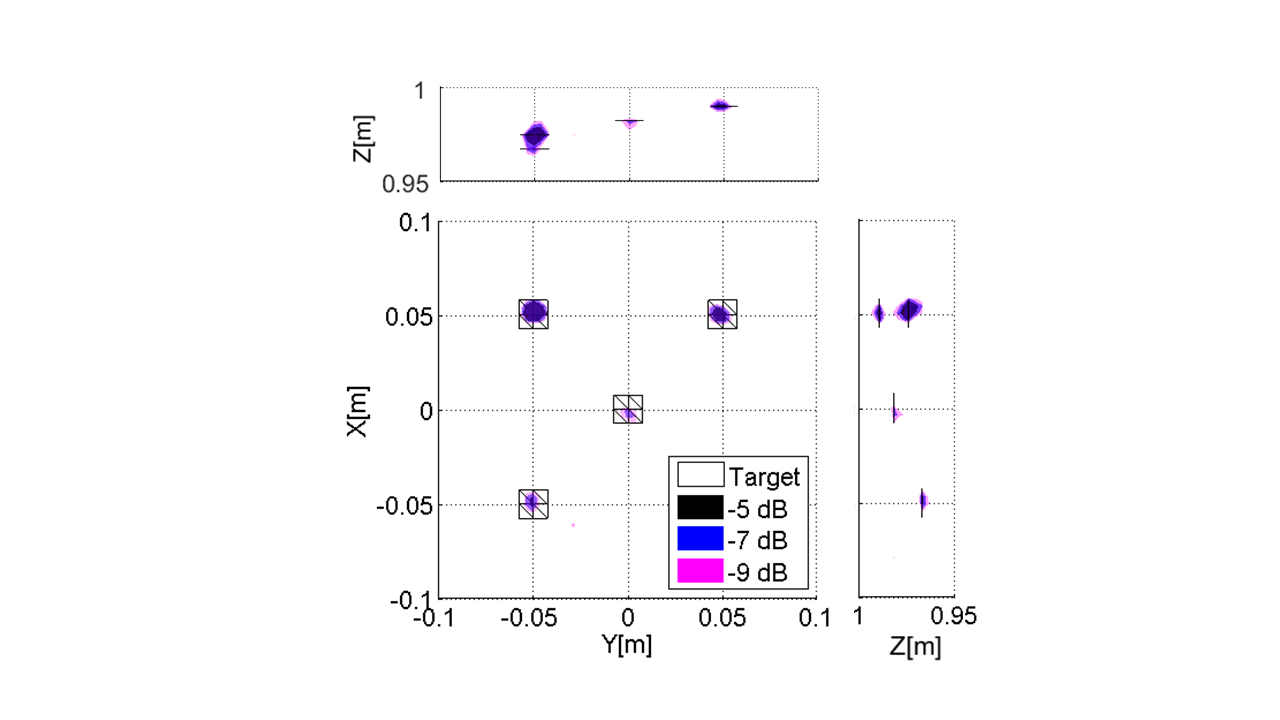}}
     \subfigure[]{
          \label{fig:CRA-ADMM}
          \includegraphics[scale=.40, trim = 72mm 10mm 30mm 15mm, clip]{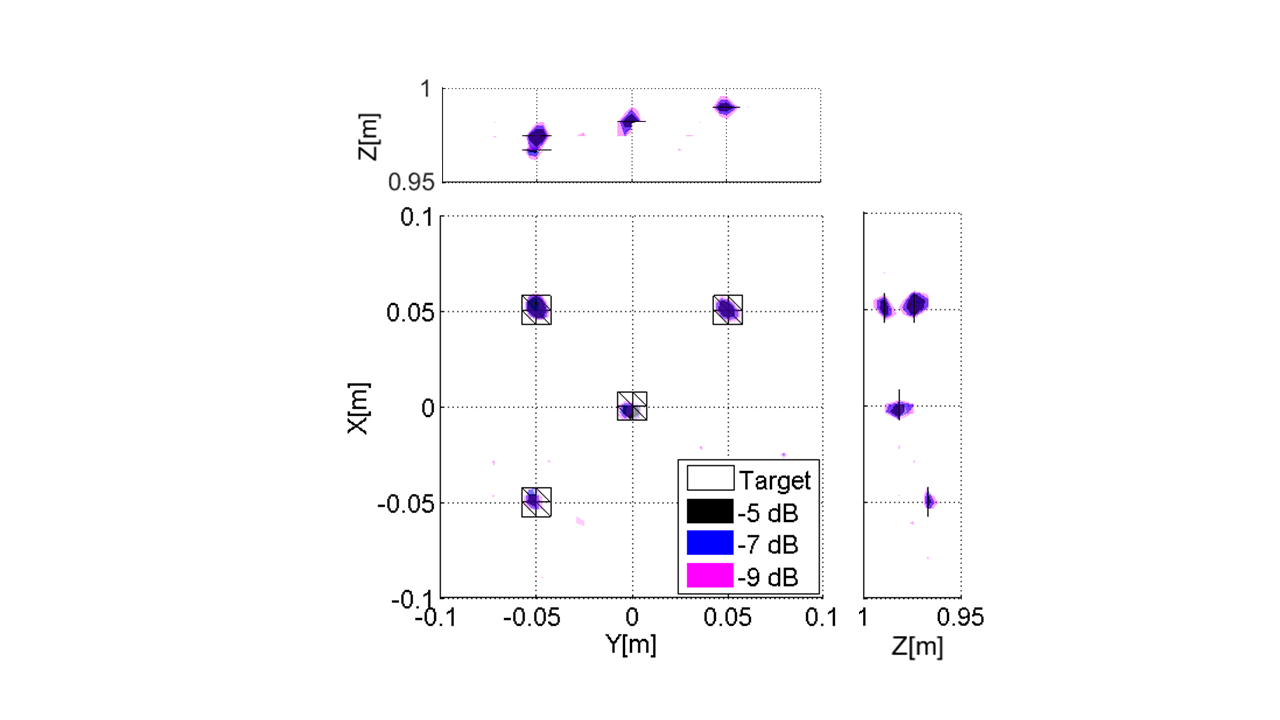}}
     \caption{Imaging reconstruction (top, front and side views) using (a) pseudoinverse, (b) NESTA, (c) ADMM. The targets are represented with the transparent black triangles
        and the reconstructed reflectivity is presented in the colored map.}
     \label{fig:Imaging}
\end{figure}

\begin{figure}[htp]
\centering
        \includegraphics[scale=.41, trim = 63mm 10mm 0mm 9mm, clip]{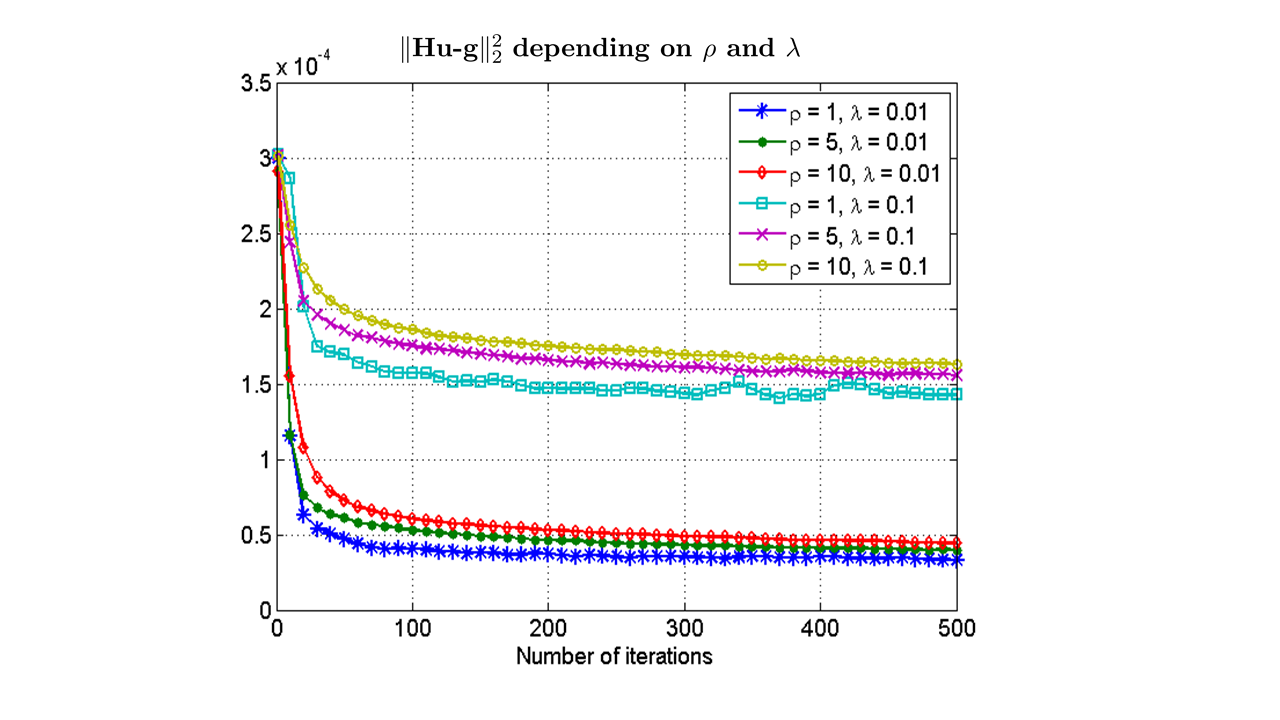}
        \caption{Convergence of ADMM solution for different parameters.}
    \label{Convergence}
\end{figure}

With the paremeters shown in Table \ref{tab:config}, the sensing
matrix $\bf{H}$ has a size of $93\times25000$. The proposed method
divides $\bf{H}$ into $N=31$ submatrices of size $3\times25000$,
which is used for each optimization of $\textbf{u}_i$. As a result
of applying the \textit{matrix inversion lemma}, only $31$
matrices of dimension $3\times3$ need to be inverted instead of a
large $25000\times25000$ matrix. The inversion of these $31$
matrices are performed just once; and they are used afterwards in
each iteration, as indicated in equation \eqref{x_solution}. The
proposed ADMM algorithm highly accelerates the optimization
process. Figure \ref{fig:Imaging} shows the imaging results using
(a) a traditional pseudo-inverse approach, where many artifacts
appear, (b) NESTA algorithm \cite{Becker2011} and (c) the ADMM
method, with a norm-1 weight of $\lambda=0.01$ and a value of
$\rho=1$, for a structure of 4 targets. Despite a few artifacts
may appear in this process, the regularized ADMM solution clearly
outperforms the pseudo-inverse solution in terms of image quality.
Additionally, the ADMM algorithm solved the problem in just
3\textit{s} for 500 iterations, while the NESTA algorithm solved
the problem in 203\textit{s}, thus showing the efficacy of the
proposed approach. In Fig. \ref{Convergence}, the ADMM convergence
process for different values of the parameters $\lambda$ and
$\rho$ is shown, including the combination used for the example in
this paper. The stability and speed of the convergence prove that
the imaging could be performed in real time.

\section{Conclusion}
This work has presented the mathematical principles of a new distributed,
consensus-based imaging algorithm using the norm-one-regularized
ADMM for a Compressive Reflector Antenna. The explanation of the whole methodology, the graphical comparison between other techniques and the convergence process have been explained in this paper. Besides the simplicity of the proposed algorithm, it outperforms both
traditional pseudo-inverse imaging algorithms, in terms of image
quality, and current state of the art iterative algorithms (i.e.
NESTA), in terms of computational cost.

\section*{ACKNOWLEDGEMENT}
This work has been funded by NOAA (NA09AANEG0080) and DHS
(2008-ST-061-ED0001).

\bibliography{references,ADMM}
\bibliographystyle{IEEEtran}

\end{document}